\documentclass[]{emulateapj_nodraft}

\usepackage{graphicx} 
\usepackage{pifont}

\shorttitle{The Magnetic Topology of CME Sources}
\shortauthors{Ugarte-Urra, Warren \& Winebarger}

\begin{document}


\title{The Magnetic Topology of Coronal Mass Ejection sources}

\author{Ignacio Ugarte-Urra\altaffilmark{1} \and Harry P. Warren}
\affil{E. O. Hulburt Center for Space Research, Code 7670, Naval
       Research Laboratory, Washington, DC 20375, USA;\\ iugarte@ssd5.nrl.navy.mil, hwarren@nrl.navy.mil}
\altaffiltext{1}{Also at College of Science, George Mason University, 4400 University Drive, Fairfax, VA 22030, USA}
\and
\author{Amy R. Winebarger}
\affil{Department of Physics, Alabama A\&M University, 4900 Meridian Street Normal, AL 35762, USA;\\ winebarger@physics.aamu.edu}


\begin{abstract}
In an attempt to test current initiation models of coronal mass ejections (CMEs), with an emphasis on
the magnetic breakout model, we inspect the magnetic topology of the sources of 26 CME events in the 
context of their chromospheric and coronal response in  
an interval of approximately nine hours around the eruption onset. First, we perform current-free
(potential) extrapolations of photospheric magnetograms to retrieve the key topological ingredients,
such as coronal magnetic null points. Then we compare the reconnection signatures observed in the
high cadence and high spatial resolution of the Transition Region And Coronal Explorer (TRACE) images
with the location of the relevant topological features. The comparison reveals that only seven events
can be interpreted in terms of the breakout model, which requires a multi-polar topology
with pre-eruption reconnection at a coronal null. We find, however, that a larger number of events 
(twelve) can not be interpreted in those terms. No magnetic null is found in six of them. Seven other 
cases remain difficult to interpret. 
We also show that there are no systematic differences between the CME speed and flare energies of
events under different interpretations. 
\end{abstract}



\section{Introduction}
Coronal Mass Ejections (CMEs) are solar eruptions that expel up to 10\,$^{16}$ g of
coronal material at speeds that range a few km\,s$^{-1}$ to over 2000 km\,s$^{-1}$ 
\citep[e.g.][]{vourlidas02a,yurchyshyn05a}. 
Though it is clear that solar magnetic fields play an important role in confinement
of plasma in the corona and the storage of free energy before a CME, developing a
detailed understanding of how the evolution of the Sun's magnetic fields can trigger
the sudden release of mass and energy has proved to be very challenging. Many
theories of coronal mass ejections have been proposed 
\citep[see reviews][]{forbes00a,klimchuk01a,zhang05a} that rely on different methods
of energy release. Models, such as the ``magnetic breakout'' model
\citep{antiochos99a,lynch04a}, the ``tether cutting'' model \citep{moore80a,sturrock89a} 
and the ``flux rope'' models \citep[e.g.][]{forbes91a,lin01a,amari00a,amari04a,torok05a} 
make specific predictions on the magnetic complexity required for an eruption and 
the time and location of the magnetic reconnection that either drives or results
from the CME.

The breakout model requires the presence of a multi-polar magnetic configuration and 
reconnection at a coronal null that allows the transfer of magnetic flux between 
flux systems. This means that 
reconnection is expected above the sheared expanding arcade, just before or at the 
eruption,  followed by the standard flare reconnection below the erupted material. 
The flux rope and tether cutting models, in contrast, do not require complex 
multi-polar fields and predict the reconnection of low-lying fields very close in 
time, just before or at, the onset of the CME. In these cases, the presence of a 
coronal null is not required, although the null can be present as a 
passive actor in the eruption.

Observationally, there is some evidence for both scenarios: eruptions in multi-polar 
active regions with an active coronal null \citep[e.g.][]{aulanier00a,sterling01a,manoharan03a,gary04a},
and eruptions in bipolar regions or even multi-polar with a passive null \citep[e.g.][]{moore01a,li06a}. 
These studies, however, were generally limited to investigating a single event
and often employed different data sets and analysis techniques.

In the present work, we perform a systematic study of a large number of events
within the same analysis framework. In the high cadence and high  spatial 
resolution images of TRACE \citep[Transition Region And Coronal Explorer,][]{handy99a}, 
we inspect the extreme ultraviolet (EUV) and ultraviolet (UV) response in a time 
range that spans approximately nine hours around the eruption time 
and analyze it in the context of potential field extrapolations of the photospheric
magnetic field. The comparison allows us to determine the presence (or absence)
of coronal magnetic nulls and their association with the UV-EUV response, which flux 
systems have an active role in the eruption, and the relevant timings for the 
event. 

Our results indicate that, even though several eruption events can be
interpreted in terms of breakout reconnection, a larger number of them do
not fulfill the requirements in terms of the magnetic topology or the timing of
the UV-EUV response. 

The paper is subdivided into several sections. In \S~\ref{data} we describe
the data selection and the different steps in the analysis, including an illustration
through two sample cases. 
\S~\ref{results} introduces a summary of the results before having a closer look at
specific properties, and finally in section~\ref{conclusions} we give our final remarks and
conclusions.


\begin{deluxetable}{rlrccrc}
\footnotesize
\tablecaption{List of CME events and their sources}
\tablewidth{0pt}
\tablerefs{$^1$\citet{zhou06a}; $^2$\citet{gopalswamy04a}; $^3$\citet{aulanier00a}}
\tablehead{
 	&   & \multicolumn{2}{c}{GOES flare}  & LASCO &  & \\
No. 	&  Date & Class & Time & CME Time 	& Source & Ref}
\startdata
1		& 1998/07/14   & M4.4  &  13:00        & \nodata&  AR8270      & $^3$\\   
2		& 1999/05/10   & M2.5  &  05:31        & 05:50 &  AR8539       & $^1$\\  
3		& 1999/07/19   & C4.2  &  02:13        & 03:06 &  AR8631       & $^1$\\  
4		& 1999/09/13   & C2.6  &  16:49        & 17:31 &  AR8693       & $^1$\\   
5		& 2000/06/06   & X2.3  &  15:25        & 15:54 &  AR9026       & $^{1,2}$ \\  
6		& 2000/06/07   & X1.3  &  15:54        & 16:30 &  AR9026       & $^1$\\  
7	 	& 2000/07/14   & X5.8  &  10:24        & 10:54 &  AR9077       & $^{1,2}$ \\  
8		& 2000/07/25   & M8.1  &  02:50        & 03:30 &  AR9097       & $^1$\\   
9		& 2000/09/06   & C2.2  &  15:29        & 16:30 &  AR9154       & $^1$\\   
10		& 2000/09/25   & M1.9  &  02:16        & 02:50 &  AR9167       & $^1$\\   
11		& 2000/11/24a  & X2.0  &  05:02        & 05:30 &  AR9236       & $^1$\\   
12		& 2000/11/24b  & X2.4  &  15:14        & 15:30 &  AR9236       & $^{1,2}$ \\  
13		& 2000/11/24c  & X1.9  &  22:00        & 22:06 &  AR9236       & $^1$\\   
14		& 2001/04/08   & C1.5  &  23:38        & 00:06 &  AR9415       & $^1$\\	    
15		& 2001/04/09   & M8.0  &  15:35        & 15:54 &  AR9415       & $^1$\\	    
16		& 2001/04/10   & X2.2  &  05:26        & 05:30 &  AR9415       & $^{1,2}$ \\  
17		& 2001/04/11   & M2.4  &  13:27        & 13:32 &  AR9415       & $^1$\\	
18		& 2001/09/22   & C2.7  &  09:25        & 09:42 &  AR9626       & $^1$\\   
19		& 2001/10/19a  & X1.6  &  01:05        & 01:27 &  AR9661       & $^1$\\   
20		& 2001/10/19b  & X1.6  &  16:31        & 16:50 &  AR9661       & $^{1,2}$ \\  
21		& 2001/10/25   & X1.4  &  15:03        & 15:26 &  AR9672       & $^1$\\   
22		& 2001/11/28   & M6.8  &  16:36        & 17:30 &  AR9715       & $^1$\\   
23		& 2001/12/13   & X6.1  &  14:30        & 14:54 &  AR9733       & $^1$\\   
24		& 2002/03/15   & M2.2  &  23:17	       & 23:06 &  AR9866       & $^2$\\   
25		& 2002/07/15a  & X3.1  &  20:07        & 20:30 &  AR10030      & $^1$\\  
26		& 2002/07/15b  & M1.8  &  20:28        & 21:30 &  AR10030      & $^2$\\  
\enddata
\label{tab:sample}
\end{deluxetable}

\begin{figure}[htbp!]
\centering
\includegraphics[bb=100 70 410 370, width=8.0cm]{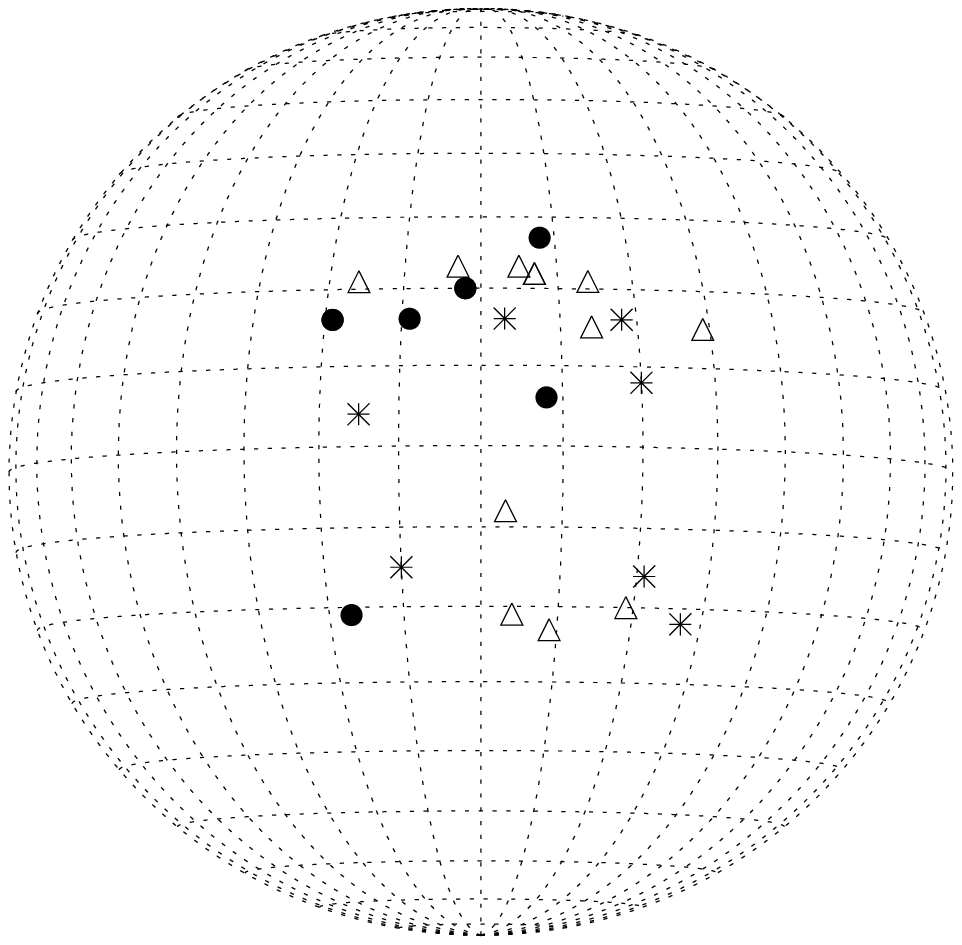}
\includegraphics[width=5cm]{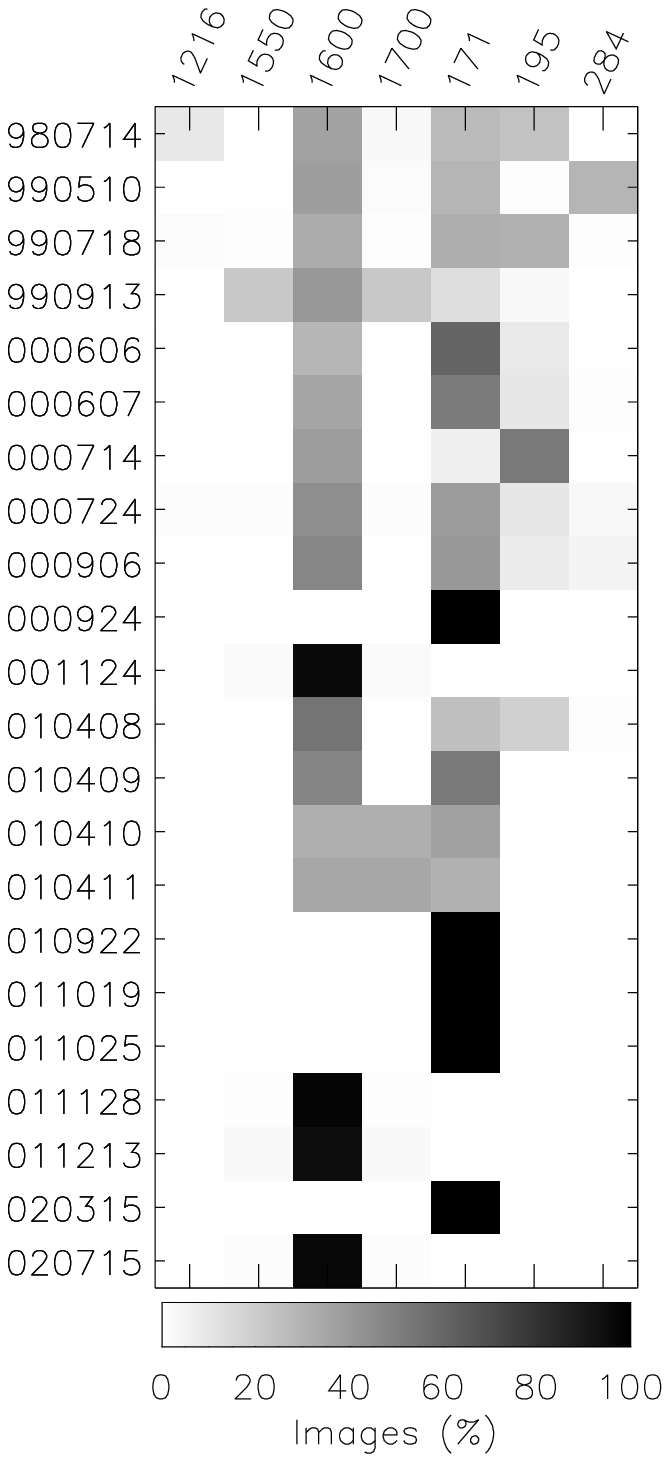}
\caption{Top: distribution of source locations. Symbols indicate different type of 
events. Full description in the text and in Table~\ref{tab:events}. Bottom: TRACE observational coverage.
Percentage of total images per date and passband.}
\label{fig:sources}
\end{figure}

\section{Data selection and analysis}
\label{data}
\subsection{Sample}
In the present study our goal is to analyze the plasma response during the 
eruptions in the context of the magnetic topology of the sources. Therefore,
we want to study CME sources that fulfill certain criteria. We look for 
identified sources of cataloged CMES; sources that have been observed in the 
UV-EUV at a minimum spatial and temporal resolution and are located within 
400\arcsec\ of disk center. These requirements are necessary to obtain reliable 
magnetic field extrapolations and a comprehensive understanding of the timings
and locations relevant to the eruption.

Our starting working sample is the one provided by \citet{zhou06a}: a total
of 288 earth-directed halo CMEs with their identified on-disk sources, observed by
LASCO \citep[Large Angle and Spectrometric Coronograph Experiment,][]{brueckner95a} 
in the interval from March 1997 to December 2003. In particular the 178 CMEs that 
lie in the interval between April 1998 to December 2001. We also worked with a
list of 60 major solar energetic particle events of cycle 23 \citep{gopalswamy04a}.  
From those two samples, we selected those events that were observed by TRACE 
in any of its UV-EUV 
passbands and were located, at the time of the eruption and observations, within 
400\arcsec\ of disk-center. We filtered the data further to get rid of cases with 
large data gaps, insufficient field of view or large and complex magnetic 
configurations that are difficult to handle with our Cartesian extrapolation code. 
In total, we selected 25 study cases, plus a well known breakout case: 
the 1998 July 14 event \citep{aulanier00a}. A summary of the observations is
presented in Table~\ref{tab:sample}.
There are 18 different sources, all active regions, and some of them 
produce multiple events. The eruptions exhibit GOES (Geostationary 
Operational Environmental Satellites) flares in the range C1.5 (1-8~\AA\ 
flux = 1.5$\times10^{-6}$ W m$^{-2}$)  to X5.8 (5.8$\times10^{-4}$ W m$^{-2}$). 

\subsection{EUV-UV data}
For each event, we inspected TRACE images in the 9 hour interval centered around
one hour before the first observed time of the CME in LASCO. The observing programs
for each event were different. The events were mostly covered with 
the 171~\AA\ or the 1600~\AA\ passband or sometimes both (see Figure~\ref{fig:sources}), 
which correspond to plasma in the lower corona ($\approx$1MK) and lower 
chromosphere, respectively. The observing cadence depends on the program,
but typical values are of the order of a minute for the EUV images and seconds
for the UV. The images were reduced using standard software and 
were co-aligned to the magnetograms by cross-correlating the magnitude of the magnetic 
flux density to the 1600~\AA\ intensities, when available. Otherwise, EIT 
\citep[Extreme-ultraviolet Imaging Telescope,][]{delaboudiniere95a} 171 \AA\ 
images were used.
\begin{figure*}[htbp!]
\centering
\includegraphics[angle=90,width=17.5cm]{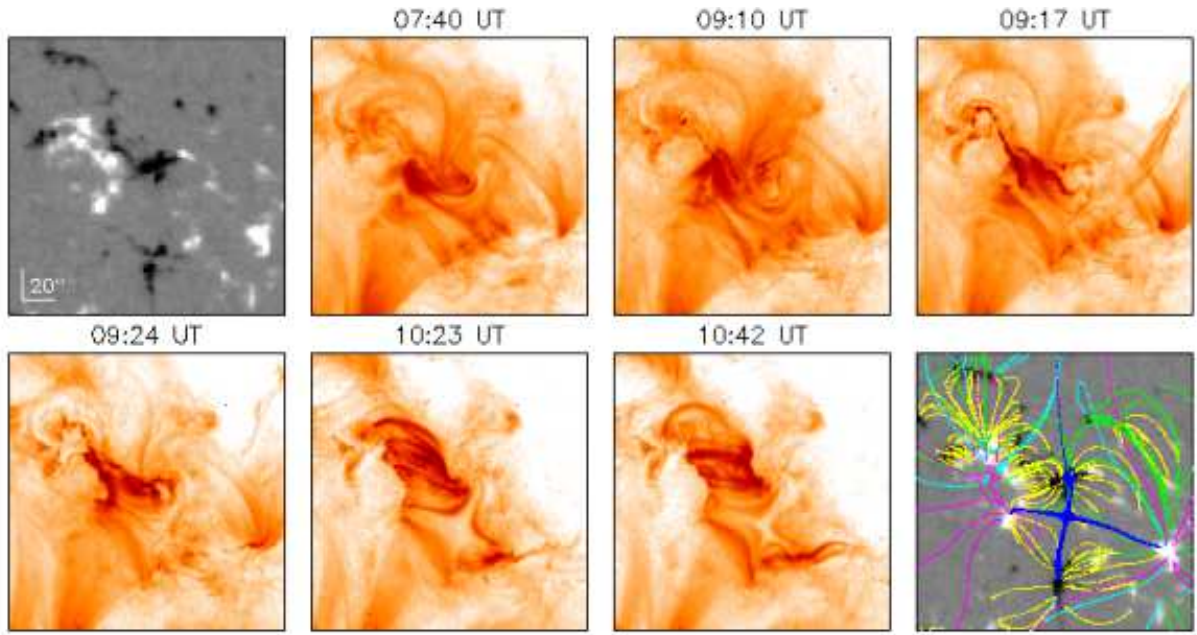}
\caption{AR9626. September 22, 2001. Top left and bottom right panels show the magnetogram and the potential extrapolation
obtained from it. Field line colors: blue are field lines around a coronal null point; other colors imply different 
field line lengths. The rest of the panels show the evolution
of the EUV plasma emission in the TRACE 171~\AA\ filter during the eruption. Notice the reverse color scale: darker means 
brighter. This figure is also available as mpeg animation in the electronic edition of the {\it Astrophysical Journal}.}
\label{fig:example1}
\end{figure*}
\begin{figure*}[htbp!]
\centering
\includegraphics[angle=90,width=17.5cm]{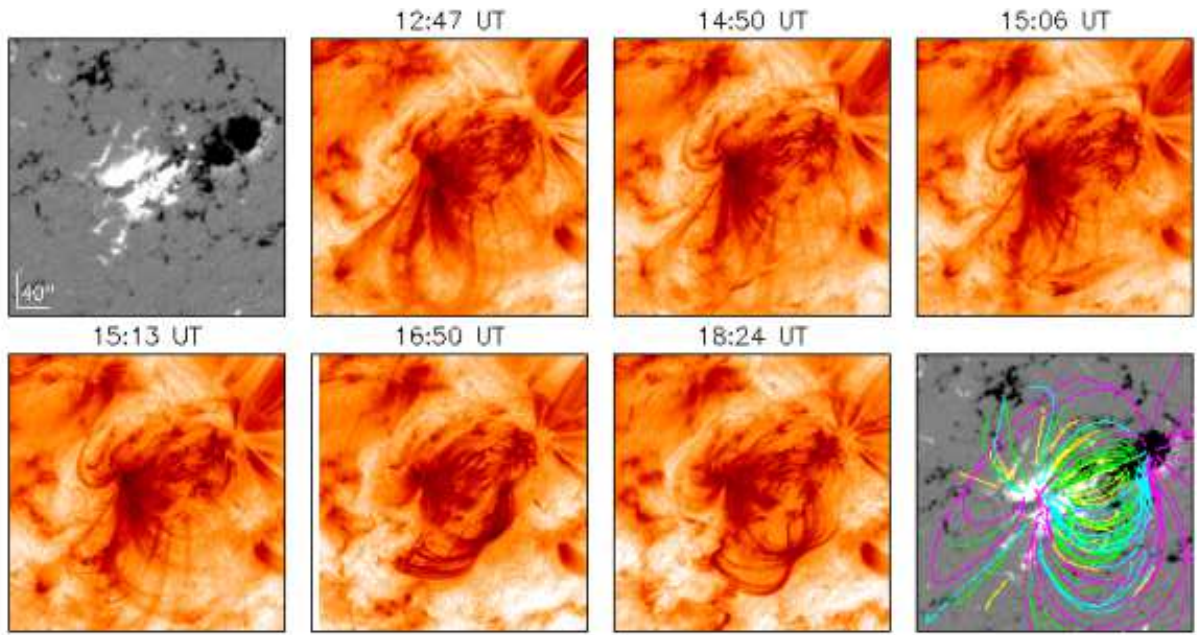}
\caption{AR9145. September 6, 2000. Top left and bottom right panels show the magnetogram and the potential extrapolation
obtained from it. The rest of the panels show the evolution of the EUV plasma emission in the TRACE 171~\AA\ filter 
during the eruption. This figure is also available as mpeg animation in the electronic edition of the {\it Astrophysical Journal}.}
\label{fig:example2}
\end{figure*}


\begin{deluxetable*}{rlccccccc|c}
\footnotesize
\tablecaption{Characterization of the events}
\tablewidth{0pt}
\tablehead{
 	&  	&     & 	& 	   & 		   & \multicolumn{2}{c}{Reconnection}	   &		  & Filament\\
No.	& Date	&  MP &  NP  	& SNL	   &   PFLR        & Location	& Timing		   &	  Model   & eruption}
\startdata
1   & 1998/07/14   & \checkmark & \checkmark & \checkmark  & Within fan 	& Fan \& spine  	   &  At		 & \textbullet  				      & $\times$ \\
2   & 1999/05/10   & \checkmark & \checkmark & \checkmark  & Within fan 	& Fan \& spine  	   &  At		 & \textbullet  				      & \checkmark \\
8   & 2000/07/25   & \checkmark & \checkmark & \checkmark  & Within fan 	& Null  		   &  At		 & \textbullet  				      & U\\
23  & 2001/12/13   & \checkmark & \checkmark & U	   & Within fan 	& Fan \& spine  	   &  U 		 & \textbullet  				      & U \\
25  & 2002/07/15a  & \checkmark & \checkmark & \checkmark  & Within fan 	& Fan \& spine  	   &  Before		 & \textbullet  				      & U \\
26  & 2002/07/15b  & \checkmark & \checkmark & \checkmark  & Within fan 	& Fan \& spine  	   &  Before		 & \textbullet  				      & U \\
18  & 2001/09/22   & \checkmark & \checkmark & \checkmark  & U (X-point)	& Null  		   &  Before		 & \textbullet  				      & \checkmark \\
    &	    	   &	        & 	     &  	   &			&			   &			 &						      &  \\
21  & 2001/10/25   & \checkmark & \checkmark & \checkmark  & Within fan 	& U			   & U  		 & \textasteriskcentered			      & U \\
4   & 1999/09/13   & \checkmark & \checkmark & \checkmark  & Within fan 	& Filament$^{\dagger}$	   & Before	         & \textasteriskcentered			  & \checkmark \\
7   & 2000/07/14   & \checkmark & \checkmark & \checkmark  & Within fan 	& NL$^{\dagger}$	   & Before	         & {\footnotesize $\triangle$}  		  & \checkmark \\
19  & 2001/10/19a  & \checkmark & \checkmark & \checkmark  & Within fan 	& U			   & U			 & \textasteriskcentered			      & U \\
15  & 2001/04/09   & \checkmark & \checkmark & \checkmark  & Within fan 	& NL$^{\dagger}$	   & U  		 & {\footnotesize $\triangle$}			      & U \\
10  & 2000/09/25   & \checkmark & \checkmark & \checkmark  & U  		& Spine?		   & At 		 & \textasteriskcentered			      & U \\
5   & 2000/06/06   & \checkmark & \checkmark & \checkmark  & U  		& NL$^{\dagger}$	   & Before$^{\dagger}$  & {\footnotesize $\triangle$}  		      & U \\
6   & 2000/06/07   & \checkmark & \checkmark & \checkmark  & U  		& NL$^{\dagger}$	   & At$^{\dagger}$	 & {\footnotesize $\triangle$}  		      & \checkmark \\
14  & 2001/04/08   & \checkmark & \checkmark & \checkmark  & U  		& U			   &	   U		 & \textasteriskcentered			      & U \\
17  & 2001/04/11   & \checkmark & \checkmark & \checkmark  & U			& NL$^{\dagger}$	   & At$^{\dagger}$	 & \textasteriskcentered 			      & U \\
3   & 1999/07/19   & \checkmark & \checkmark & \checkmark  & Outside fan	& Filament$^{\dagger}$/PFLR& At 		 & {\footnotesize $\triangle$}  		      & \checkmark \\
24  & 2002/03/15   & \checkmark & \checkmark & \checkmark  & Outside fan 	& PFLR			   & After		 & {\footnotesize $\triangle$}  		      & U \\
    &	    	   &	        & 	     &  	   &			&			   &			 &						      &  \\
11  & 2000/11/24a  & \checkmark & $\times$  & U	   &			& NL$^{\dagger}$	   & U  		 & {\footnotesize $\triangle$}   		      & U \\
12  & 2000/11/24b  & \checkmark & $\times$  & U 	  &		       & NL$^{\dagger}$ 	  & Before$^{\dagger}$  & {\footnotesize $\triangle$}			     & U \\
13  & 2000/11/24c  & \checkmark & $\times$  & U 	  &		       & NL$^{\dagger}$ 	  & Before$^{\dagger}$  & {\footnotesize $\triangle$}			     & U \\
16  & 2001/04/10   & \checkmark & $\times$  & \checkmark  &		       & Filament$^{\dagger}$	  & At$^{\dagger}$	& {\footnotesize $\triangle$}			     & \checkmark \\
20  & 2001/10/19b  & \checkmark & $\times$  & \checkmark  &		       & NL			  &	 Before 	& {\footnotesize $\triangle$}			     & \checkmark \\
22  & 2001/11/28   & \checkmark & $\times$  & U 	  &		       & NL$^{\dagger}$ 	  & Before$^{\dagger}$  & \textasteriskcentered 			     & U \\
    &	    	   &	        & 	     &  	   &			&			   &			 &						      &  \\
9   & 2000/09/06   & $\times$   & $\times$  & \checkmark  &			& NL$^{\dagger}$	   & Before$^{\dagger}$  & {\footnotesize $\triangle$}  		      & \checkmark \\

\enddata
\tablecomments{MP: multipolar; NP: null point; SNL: sheared neutral line; PFLR: post-flare loop ribbons; U: unclear; NL: neutral line; $\bullet$: consistent with the 
		topology and observational predicitions of the breakout model \citep{antiochos99a}; $\vartriangle$: inconsistent with that model; 
		$\ast$: remains unclear; $^{\dagger}$: intensity enhancement, but not necessarily due to reconnection.}
\label{tab:events}
\end{deluxetable*}


\subsection{Magnetic field extrapolation}
In order to examine the magnetic field topology of the sources, we performed
potential field extrapolations, the current free case of a force-free field
approximation. The method and solutions, obtained by means of Fourier 
transforms in a Cartesian coordinate system, can be found in 
\citet{alissandrakis81a} and \citet{gary89a}. As boundary conditions we used 
the line-of-sight component of the photospheric magnetic field, as given by 
full disk magnetograms from MDI \citep[Michelson Doppler Imager,][]{scherrer95a} 
for all the cases except for the July 14 1998 event in which we used a Kitt Peak
magnetogram. 

For each source, we selected a square field of view with sizes ranging 
between 350 and 480 Mm long, depending on the area occupied by the dominant 
active region and neighbouring flux systems. In cases where several 
active regions are found in close proximity, a reduced field of view can result
in too localized connectivities that do not represent the interconnectivity
between different regions. For those cases, several fields of view were
tested, trying to reproduce the connectivities provided by a potential field 
source-surface model \citep[PFSS,][]{schrijver03a}, a full Sun potential 
extrapolation in spherical coordinates. This model 
provides a better context in terms of the surrounding flux that can affect
the flux systems that we are interested in. On the other hand, the resolution
at smaller scales is insufficient to deal with finer structures within a single
active region, which is crucial for a study like ours.

Once the full 3D magnetic field distribution is calculated, we proceed to 
locate coronal null points, i.e. locations where the magnetic field vanishes. 
Our code essentially looks for changes in the 
direction of the magnetic field vectors surrounding each point. When large 
changes in the direction of the magnetic field are found, namely a minimum in 
the dot product of the field vectors of two adjacent positions, we then search 
for nearby minima as well as for minima along field lines that pass near 
that point.

It is important to note that erupting systems are generally
non-potential \citep[e.g.][]{schrijver05a}.
The excess of magnetic energy is what allows the system to relax in a violent 
way and expel coronal mass. Therefore, we do not expect to reproduce the observed 
loops in detail. The non-potentiality, however, can be localized and confined 
to one topological domain where the field is strongly sheared. For the large scale 
topology, potentiality can be a fair approximation \citep[e.g.][]{aulanier00a,li06a} of 
the relevant topological domains and the presence of key 
topological ingredients like coronal null points. Null points are stable topological 
features and cannot be easily destroyed \citep{greene88a}. A change in magnetic shear 
can result in a shift of its position \citep{demoulin94a}, its height for example. 
We have confirmed this expected general agreement by visually comparing the
loop connections in the EUV images to the flux systems result of 
the extrapolation. Nulls outlined in the coronal images are also found in the
extrapolations, sometimes slightly shifted from their predicted position.

\subsection{Analysis: Two Sample Cases}
\label{analysis}
We will illustrate our strategy through two representative 
examples of the events in the sample. We begin with event 18 in Table~\ref{tab:sample}.
The CME associated with this event was first seen in LASCO at 09:42 UT on September 22, 
2001. A weak C2.7 GOES flare peaking at 09:25 UT was associated with it. The source is 
active region AR9626 which can be seen in detail in Figure~\ref{fig:example1}.

Figure~\ref{fig:example1} shows the magnetic topology and the 171 \AA\ plasma evolution during the eruption. 
The magnetogram on the top left corner shows in its center a leading negative 
(black) polarity that in the EUV images is connected to the trailing positive polarity 
by sheared EUV loops. The magnetic shear is obvious from the comparison of the 
loops across the active region neutral line (NL), seen in the 07:40 UT pre-eruption 
171\AA\ image, to the field line connectivities across that same NL in the potential 
extrapolation of the bottom right panel. The extrapolation shows the presence of a 
coronal null point, in the shape of an 
X (blue lines), between the edges of AR9626 and some neighbouring flux. 

A close 
inspection of the pre-eruption NL shows highly sheared loops and absorption EUV 
features, typical of prominence material, from the beginning of the observational 
data, which in this particular case is as early as 8 hours before the eruption. The 
NL loops, as they slowly evolve, begin to reveal the presence of the null by 
outlining the top of the X (07:40 UT panel). At around 08:59 UT the expansion of 
the inner sheared loops becomes evident (compare with the 
09:10 UT panel) and the interaction with the coronal null at around 09:13 UT produces 
the violent eruption (09:17 UT panel) that results in the disappearance of the outer 
sheared arcade and the subsequent formation of the post-flare loops (see 09:24 UT 
through 10:42 UT). This final and relaxed configuration is close to potential. 
Some of the post-eruption loops outline the null point location.

According to our introductory considerations, this is an example of breakout 
eruption candidate. There is an evident coronal null that becomes active before the eruption
and allows reconnection to take place and open the way out to the erupting material. 

A totally different case is event 9. The CME associated to this event was first seen in 
LASCO at 16:30 UT on September 6, 
2000. A weak C2.2 GOES flare peaking at 15:29 UT was associated with it. The source is 
active region AR9145 and can be seen in detail in Figure~\ref{fig:example2}. Like 
Figure~\ref{fig:example1}, it shows the magnetic topology and the 171 \AA\ plasma 
evolution during the eruption. In this case, however, the source is fundamentally bipolar 
and there is no coronal null present in the extrapolated field. As a result, with the 
current analysis, the eruption can not be interpreted in terms of breakout. 

At the larger scales, the magnetic field topology of the active region is close to 
potential even before the eruption. The large overlying loops in the 12:47 UT TRACE
image show connections that are similar to the field lines in the bottom right panel. 
Under this arcade of loops, we find sheared structures and prominence material that
appears suspended in a plane almost perpendicular to the plane of the arcade. At
around 13:41 UT there is a flaring event close to the NL. Minutes later, the 
prominence material starts to 
slowly rise. In its rising phase it interacts with the overlying arcade
that gets pushed aside, starting at around 14:50 UT, while the to be-erupted 
material makes its way up increasing its speed. Frames 15:06 UT and 15:13 UT
represent that phase. The erupted plasma is seen in absorption (lighter in these
reverse scale images) near the bottom of the 14:50 UT still. It is seen as a bright
(dark in the image) feature in the 15:06 UT and reaches the edge of the field of
view in the 15:13 UT frame. Finally, we see the formation of the post-flare
loops outlining the potential topology of the bipolar source (frames at 16:50 UT
and 18:24 UT). 

Since this event clearly lacks a relevant null point it does not
support the breakout reconnection model.
 

\section{Results and discussion}
\label{results}
The results of our analysis (\S~\ref{analysis}) of each of the 26 events presented in
Table~\ref{tab:sample} are discussed in this section. First we summarize the results 
in a Table of properties, then we examine them making special emphasis in important 
aspects like morphology and timing. Finally, we will discuss the CME and flare 
properties of the events.

\subsection{Summary}
The results are summarized in Table~\ref{tab:events}. In the 
table we present a comprehensive description of properties for all the
events. This is a modified version of an original table proposed in the
SHINE (Solar, Heliospheric and INterplanetary Environment) meeting of 2000, 
as a guide to differentiate observationally between different CME models.
In an attempt to classify the events by their topological properties that 
can help discriminate between current CME initiation models, our results 
remain mostly qualitative, as are the fundamental differences between these models.

In the first two columns we identify the event with a number and
its date that relates them to Table~\ref{tab:sample}. The third and forth 
state if the source is multi-polar (MP) or not, and whether the erupting 
flux system is related to a coronal null point (NP) or not. 

The next column (SNL) refers to the shear of the magnetic field near the neutral line: 
is it sheared with respect to a potential field? This is determined by visual 
inspection of TRACE movies with overlays of field lines. The field is
considered sheared if the first post-flare loops are inclined 
with respect to the orientation of the field lines across the magnetic inversion
line in the current-free model. 

The sixth column (PFLR) describes the location of post-eruption ribbons with respect 
to the null topology (if present) and is discussed in more detail in \S~\ref{morphtime}. 

The next two columns characterize the reconnection signatures in terms of location and 
timing with respect to the eruption. By reconnection signatures we mean a 
sudden EUV or UV intensity increase at the source around the time of 
the eruption. We use that term for simplicity and a clearer presentation. 
It is a matter of discussion if some of these signatures are just sudden 
energy depositions or density enhancements of a different origin. We 
assume that these brightenings are signatures of reconnection when they outline 
a topology that favours the reconnection, i.e. fan and spine \citep{lau90a,priest96a} 
or classical post-flare loop reconnection, and we label the rest with 
$^{\dagger}$ to stress that the origin could be unrelated to the reconnection of 
field lines. Details for each event are discussed in the appendix. 
Post-flare loops are seen in every case.

The next to the last column (Model) indicates our interpretation of the event. 
The symbol $\bullet$ indicates 
that the event is consistent with the topology and observational predictions 
of the breakout model \citep{antiochos99a}. An event is included in this 
category if, before or at the time of the eruption, there are signatures 
that reveal the involvement of a coronal null, 
namely nearly simultaneous intensity enhancements at the location of the 
footpoints of fan and spine loops or an obvious null activity like in AR9626.
Sometimes, even if there is reconnection at the null before the eruption, it 
can be preceded by activity at the magnetic inversion line. For now, we will 
consider this fact as a property that needs further investigation, and will 
classify the events with null reconnection before the eruption as breakout cases.
Further discussion about the timing is given in \S~\ref{morphtime}.

We label with a $\vartriangle$ those events whose properties seem 
inconsistent with the breakout model. It can be due to the fact that a 
coronal null, the required condition, is not present, or if present 
it simply does not play a role. Finally, $\ast$ corresponds to those cases that, 
after the analysis, remain unclear in the interpretation.

Filaments are often associated with a CME, so in the last 
column we have also considered their association to these events.
In the majority of the events we see absorptions features in EUV 
aligned with the neutral lines. In many cases it is unclear if the filament
material gets expelled during the eruption. We have assigned a
check mark to those cases in which we see considerable upward 
motion of absorption material.

There are several interesting results to be extracted from the table. 
First, all the sources but one can be considered multi-polar and 
the majority have sheared magnetic fields near the neutral line. The  
remaining cases are uncertain because we only have 1600 \AA\ images and 
we can not see 
the loops. This common pattern gets broken once we consider the rest of the
properties. A coronal null point related to flux systems involved in the 
eruption, is found for nineteen events (73\%) and in twelve of those cases the 
null's role in the eruption seems irrelevant or unclear with the current data.
Overall, we end up with seven (27\%) that are consistent with breakout,
twelve that are not (46\%) and seven (27\%) that remain uncertain under the 
current set of observations.

\begin{figure*}[htbp!]
\centering
\includegraphics[width=18cm]{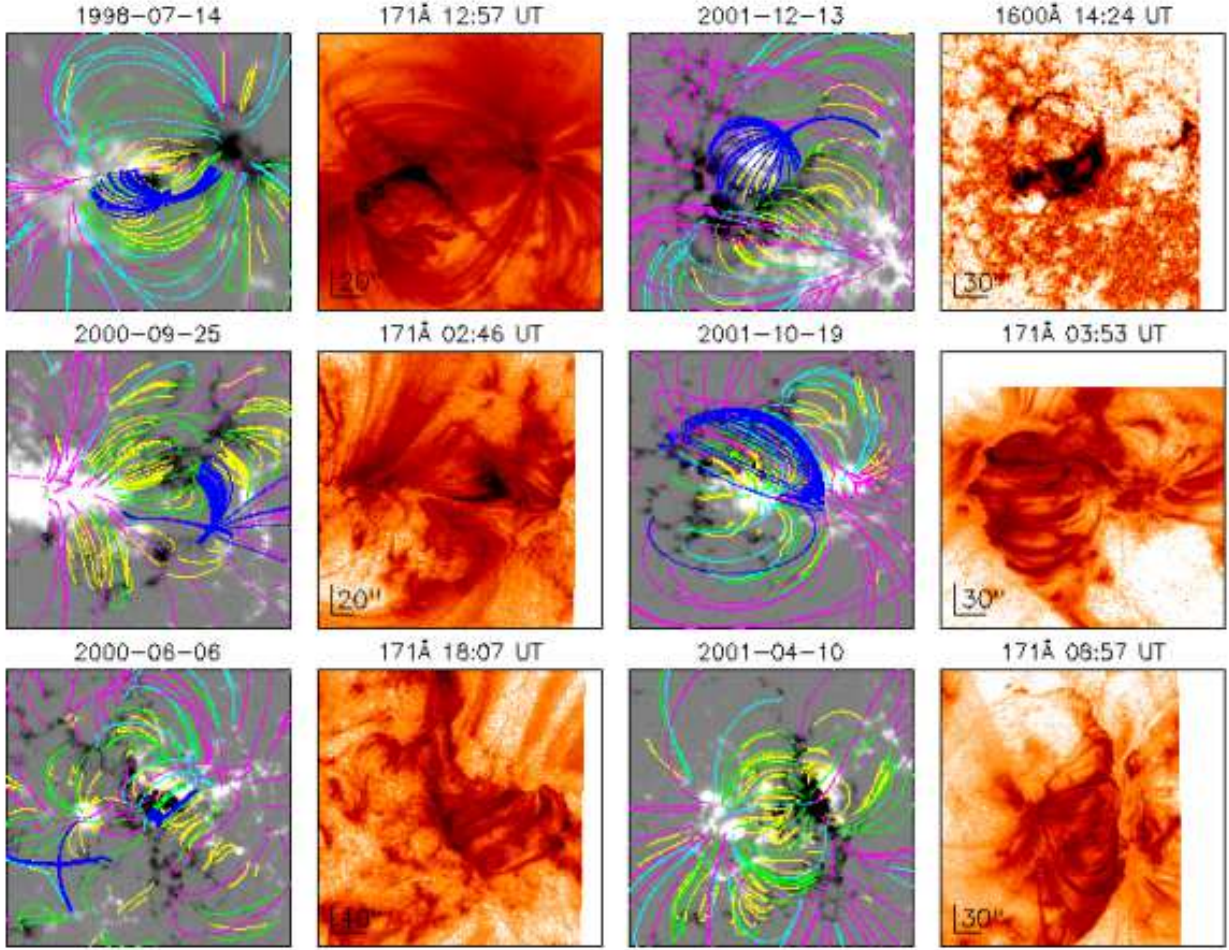}
\caption{Topology and TRACE EUV response for six CME events: two breakout candidates (top row), two non
breakout cases (bottom row) and two cases difficult to interpret (middle row). Same coding 
as previous figures. All the extrapolations are potential, except for the 2001 December 13 event (top
right) that is linear force free with $\alpha=0.005~$Mm$\,^{-1}$.
}
\label{fig:loop_vs_extrap}
\end{figure*}

\subsection{A closer look: morphology and timing}
\label{morphtime}
The ordering of the events in Table~\ref{tab:events} is prompted by grouping
their shared properties.
It is obvious that the multi-polarity (MP) and shear (SNL) are properties shared
by most of the events. The absence of a coronal null groups seven
of them in the bottom part, ruling them out from being considered as breakout 
cases. Event 22 is left uncertain for the reasons given in the appendix. 

The location of the footpoints of the post-flare loops (PFLR), i.e. the 
TRACE 1600 \AA\ ribbons, with respect to the location of the intersection of the
fan surface with the photosphere, is another important characteristic of these
events. Six of the breakout cases involve a flux system that is contained by 
the separatrix dome or fan surface originated at the null. Therefore the 
post-flare ribbons evolve {\it within the fan} in all these cases. Only one of 
them, event 18 depicted in Figure~\ref{fig:example1}, shows a different 
configuration, a 3-D X-point with no clear fan and spine within the extrapolation 
box due to the topological symmetry of the four dominant sources. 
The development of the ribbons within the fan, although characteristic of 
the breakouts events, is not exclusive of them. 
It is also seen in two non-breakout cases, events 7 and 15, and three more that 
remain uncertain. In these cases, there is a coronal null, but there are no obvious
signatures of activity there prior to the eruption.

In terms of the reconnection signatures, the table shows that those events 
interpreted as breakout exhibit intensity enhancements related directly or 
indirectly, via the fan and spine, to the null. Figure~\ref{fig:loop_vs_extrap} 
shows a more compact version of Figure~\ref{fig:example1} for six events. The top 
two panels show a brightening at the footpoints of loops that outline the spine and 
fan plus the flare ribbons, characteristic of the breakout cases in our study.
Timing should be a constraint to decide whether this null activity is a
trigger or a consequence of the eruption. To be conclusive about this aspect, 
however, is challenging. Even if the null signatures precede the eruption, 
defined as the moment in which the fields open up and the material gets 
expelled from the field of view, the fan and spine activity can be sometimes
seen preceded by an intensity enhancement near the NL and the delay can be
just a single frame in the EUV, i.e. one minute. The origin is not clear with the
current set of observations. This is observed in the two examples
in Figure~\ref{fig:loop_vs_extrap}:
the July 14, 1998 event (event 1), a breakout example analyzed in detail by 
\citet{aulanier00a}, and the December 13, 2001 case (event 23).

Those events that we do not consider candidates for breakout exhibit 
pre-eruption activity that is unrelated to any identified coronal null. The 
reconnection signatures or intensity enhancements are observed near the 
neutral line. The most likely interpretation for some of these intensity 
enhancements is the standard post-flare reconnection in the form of 
brightenings at the footpoints of reconnected small and low-lying loops. 
Other seem to be aligned with structures along the neutral line, which 
in a few cases can be directly associated to the filament. 
Among the events difficult to interpret, we find brightenings at the
neutral line and the filament, unclear interpretations too, but also an 
active null with an unresolved eruption path (event 10). In terms of timing,
it still remains challenging in some cases to state if the brightenings are
a trigger or a consequence of the eruption.

Therefore, we conclude that the morphology and the location of the energy 
deposition can be given with enough confidence as to diagnose the null's 
involvement. Timing, however, remains an unresolved issue in some of the 
eruptions. Observations with a higher temporal cadence and a complementary 
side view, which will be provided by future missions, should help to resolve 
some of these uncertainties.

The middle panels in Figure~\ref{fig:loop_vs_extrap} show two events that 
remain uncertain in our interpretation. Even though both show the presence of a
coronal null in the extrapolation, a 9 minute data gap plus evident neutral 
line activity in the right most case (event 19) and an uncertain origin of 
the eruption in the left case (event 10), leave the two cases open to 
interpretation. Five other cases in the list are left with an uncertain 
interpretation due to several considerations that are summarized case by 
case in the appendix.
 
The bottom two panels of  Figure~\ref{fig:loop_vs_extrap} show two representative
events, where breakout is not considered the most likely interpretation. In the 
right most case (event 16), we do not find a coronal null and the first signs of activity 
are seen at the filament. The bottom left case (event 5), shows no convincing activity
at the coronal nulls, but mainly at a neutral line that is extended and not associated
with a null point that would explain the magnetic flux transfer necessary for a breakout
eruption.

Finally, two general remarks with respect to the topology: first, post-flare 
loops are consistent with the potential field connectivities in the erupting 
magnetic domain, expected as a result of the relaxation of the field after the 
eruption (see Figure~\ref{fig:example1}, \ref{fig:example2} and 
\ref{fig:loop_vs_extrap}); secondly and more related to the EUV evolution, the 
ribbons migrate outward from the neutral line until they reach the location 
where the (quasi) separatrices, that encompass that particular magnetic domain, 
intersect the photosphere, i.e. where and when the post-flare reconnection ends in 
the erupting magnetic domain. The outward migration is seen for 21 (81\%) cases while five 
cases remain unclear for observational constraints or complexity in the evolution. In two
events (25 and 26) the outward migration is preceded by an inward movement toward the 
neutral line. In four events (2, 11, 12 and 13) the outward migration seen in the inner 
ribbons, close to the neutral line, is accompanied by an inward movement of outward ribbons.

\begin{figure}[htbp!]
\centering
\includegraphics[width=8.1cm]{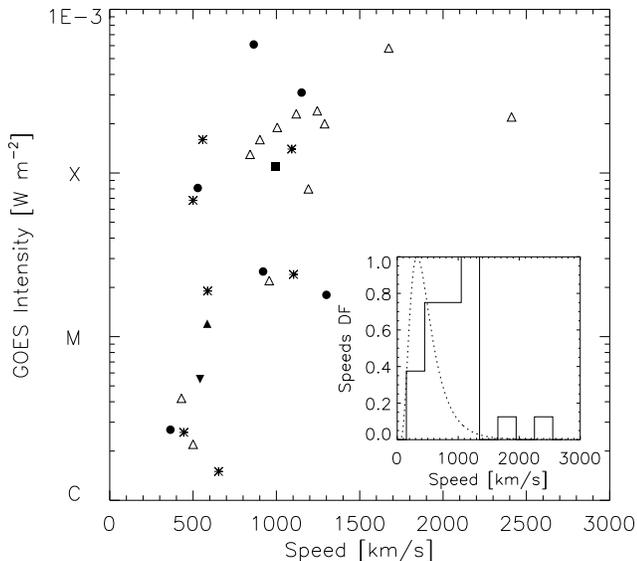}
\caption{Relationship between the GOES X-ray intensity and the CME speed. Symbols are the 
same as in Table~\ref{tab:events}, plus three extra breakout cases from 
\citet[][{\scriptsize$\blacktriangle$}, {\scriptsize$\blacktriangledown$}]{sterling01a} 
and \citet[][{\tiny $\blacksquare$}{\small}]{manoharan03a}. 
Inset: the distribution function of speeds for the paper's sample (solid line) is also 
compared to the log-normal distribution (dotted line) found by \citet{yurchyshyn05a}.}
\label{fig:speed_model}
\end{figure}

\subsection{Speeds and flare fluxes}
One question that we can ask is this: is there any difference between the CMEs 
produced by the events with different interpretations? 
In Figure~\ref{fig:speed_model} we present in data points 
the relationship between the CME plane-of-sky speed (linear fit in LASCO 
images) and the GOES X-ray intensity. The different symbols have the
same coding as in Table~\ref{tab:events}, with the addition of three
breakout candidates from the literature (filled square and filled
triangles). The correlation (Spearman's coefficient of 0.60) between 
both quantities is known \citep[][]{moon02a}. This relationship 
has also been shown in the past in terms of the kinetic energy, instead 
of the speed. For the kinetic energy, we find a correlation of 0.59, 
very close to the 0.54 presented by \citet[][see also \citealt{burkepile04a}]{hundhausen97a}.
In the context of the different models of CME initiation, it is 
important to point out that our results show no distinction in terms of 
energies or speeds between those events interpreted as breakout 
and the rest. Different observational signatures do not result in
CMEs with different energetic characteristics. 

In Figure~\ref{fig:speed_model} we also show the distribution function
of speeds in our dataset (dashed line) in comparison to the log-normal
distribution found by \citet{yurchyshyn05a} for 4315 CMEs (dotted line). 
It clearly shows that the CME events observed by TRACE are mostly in the 
higher speeds tail of the distribution. The reason behind it is 
that TRACE observations are biased towards large flaring active regions
which are often associated with fast CMEs 
\citep[see][and references therein]{burkepile04a}. Furthermore, halo
CMEs, predominant in the sample, tend to be high speed CMEs \citep{lara06a}.  
This has to be taken into consideration when extracting conclusions 
about the whole population of CMEs from a TRACE sample.


\section{Conclusions}
\label{conclusions}
We have presented results from the study of 26 CME initiation events, that
we have classified according to the two dominant schools of thought in CME 
initiation: CMEs initiated in complex topologies that require the presence of 
active coronal nulls, and CMEs that only require sheared fields in bipolar sources. 
We have found several examples that fit in the first scenario, but do not 
rule out the second. And most importantly, we have found a larger number of cases 
that do not seem to fulfill the strict requirements of the breakout model.

Our results also show that there are no systematic differences in terms of the 
energetics of the CME and the flare between events with different observational
signatures. This, in principle, rules out a scenario with two different 
mechanisms operating in different energetic domains and leading to the formation 
of distinct types of CMEs. If several models of eruption are viable, they have 
to predict a similar energetic response from the CME. The alternative 
is that the mechanism is unique and the different observational 
signatures are just a reflection of a local topology.

The results are suggestive, but not conclusive. There are limitations in data
(field of view, temperature coverage, projections, sampling) and techniques 
(oversimplified magnetic modeling) that should be addressed by future 
missions and studies.
From the observational point of view, simultaneous coverage in 1600\AA\ and 
171\AA\ has proved to be sufficient for this type of diagnostic, 
as it allows inspecting the evolution of the loops and the main footpoint 
activity. We should, however, not underestimate the limitations of interpreting
the evolution of a 3-dimensional structure in a 2-dimensional image, mainly
when that structure is developing in the direction that we do not have 
information about. Simultaneous side views, like the ones that STEREO will 
provide in the future, should help us to deal with some of the uncertainties in our
study, inherent to the lack of information in the radial direction.


\acknowledgments

This work was supported by NASA's Living with a Star Program.
We would like to thank Spiro Antiochos and Jim Klimchuk for 
many helpful discussions on this topic. TRACE is a mission of 
the Stanford-Lockheed Institute for Space Research, and part 
of the NASA Small Explorer program. SOHO is a project of 
international cooperation between ESA and NASA.

\appendix
Here we present some additional information on several of the events. It is
not meant as a detailed description of each case, but as complementary 
information for those cases that deserve some extra comments outside 
the concise characterization of Table~\ref{tab:events}. Movies for all events
can be accessed by contacting I. Ugarte-Urra.\\

\subsection{1999/05/10}
Cadence at the time of the eruption is of the order of 10 minutes, which
limitates the conclusions. In the potential field extrapolation we do not
find a point where the magnetic field completely vanishes, although we 
find a local minimum. Nevertheless from our inspection of the topology and the 
coronal images, we do believe that the coronal null is present.
\subsection{1999/07/19}
There are no signatures that relate the null to the eruption before it 
happens. The first intensity enhancements are almost simultaneously at
the filament and the ribbons of post-flare loops, which are located 
under the spine, outside the fan.
\subsection{1999/09/13}
Using a high resolution MDI magnetogram as a boundary condition for the
extrapolation we find a coronal null on top of the separatrix dome (fan) that
encompasses the erupting flux system. Nevertheless, the first signatures of the
eruption are seen as intensity enhancements at the filament that lies along the
magnetic inversion line. The removal of overlying flux, very clear in this
example, follows as the eruption develops.
\subsection{2000/09/25}
A coronal null is certainly associated with part of the post-flare loop formation
and most likely to the first stages in the flare. There is no clear evidence, 
however, of plasma being ejected from the area, even though the EIT movie shows
signatures of a global disturbance originated from that location. 
\subsection{2001/04/08}
The CME source is correctly identified and the images show some of the typical
ingredients in a classical eruption: opening of overlying fields, post-flare 
loop arcade. The interpretation, however, is left uncertain because the topological
association between the pre-eruption activity and the erupting arcade is unclear.
\subsection{2001/04/09}
No signatures of null reconnection can be seen in this case. The low lying 
null gets outlined by post-flare loops, but its role in the first 
stages of the eruption is not clear. Furthermore, some of the flux systems 
associated with the null are too small to allow significant flux exchange.
\subsection{2001/04/11}
To the east side of the erupting neutral line there is an evident coronal null,
both in the images and the extrapolation, that shows signatures of reconnection
before the event. Nevertheless, at the time of the eruption the sudden intensity 
enhancements are seen at the neutral line and the footpoints of post flare
loops.
\subsection{2001/10/19a}
There is a 9 minutes data gap at 00:50 UT, right at the moment the eruption
takes place. Post-flare loops outline the separatrix dome associated with a
low-lying coronal null in the west side of the erupting neutral line. Loops on
the neighbourhood of the null seem to be dragged to it before the eruption, but
there is no conclusive evidence of its role. Just before 00:50 UT, there is 
prominence material evolution and energy deposition at the neutral line.
\subsection{2001/10/19b}
The extrapolation code does not find a coronal point. We do however see
some activity near the neutral line that could be related to a negative 
parasitic polarity embedded in positive flux. It seems likely that 
a low lying null is present at that location, but it is not found due to 
poor resolution. Our conclusions do not change either way.
\subsection{2001/10/25}
There is a 20 minute data gap at the time of the eruption that makes it 
difficult to extract conclusions about the reconnection signatures: 
timing and location. The topology is compatible with breakout.
\subsection{2001/11/28}
Only UV images are available. The images show continuous pre-eruption
brightenings in several locations of the active region. During and after
the eruption the strongest intensity enhancements are constrained to a 
limited area. This area is associated with the topological domain that contains
flux connections between the leading positive polarity and a parasitic negative
flux concentration. The configuration resembles some of the breakout cases,
including a separatrix dome, however no coronal null is found in the potential 
extrapolation. The point with minimum field strength is very close to the bottom 
of the extrapolation box. 
\subsection{2002/03/15}
At the core of the active region there is negative polarity surrounded
almost completely by positive flux and trailing a leading negative 
polarity. A fan-spine configuration is found and activity is seen around
the null with loops outlining the separatrix dome. Yet, this happens
only after the eruption, being the main eruptive region the neutral line between
the two dominant flux systems outside the fan. Reconnection at the null
appears to be a consequence of the eruption.



\begin{thebibliography}{38}
\expandafter\ifx\csname natexlab\endcsname\relax\def\natexlab#1{#1}\fi

\bibitem[{{Alissandrakis}(1981)}]{alissandrakis81a}
{Alissandrakis}, C.~E. 1981, \aap, 100, 197

\bibitem[{{Amari} {et~al.}(2004){Amari}, {Luciani}, \& {Aly}}]{amari04a}
{Amari}, T., {Luciani}, J.~F., \& {Aly}, J.~J. 2004, \apjl, 615, L165

\bibitem[{{Amari} {et~al.}(2000){Amari}, {Luciani}, {Mikic}, \&
  {Linker}}]{amari00a}
{Amari}, T., {Luciani}, J.~F., {Mikic}, Z., \& {Linker}, J. 2000, \apjl, 529,
  L49

\bibitem[{{Antiochos} {et~al.}(1999){Antiochos}, {DeVore}, \&
  {Klimchuk}}]{antiochos99a}
{Antiochos}, S.~K., {DeVore}, C.~R., \& {Klimchuk}, J.~A. 1999, \apj, 510, 485

\bibitem[{{Aulanier} {et~al.}(2000){Aulanier}, {DeLuca}, {Antiochos},
  {McMullen}, \& {Golub}}]{aulanier00a}
{Aulanier}, G., {DeLuca}, E.~E., {Antiochos}, S.~K., {McMullen}, R.~A., \&
  {Golub}, L. 2000, \apj, 540, 1126

\bibitem[{{Brueckner} {et~al.}(1995){Brueckner}, {Howard}, {Koomen},
  {Korendyke}, {Michels}, {Moses}, {Socker}, {Dere}, {Lamy}, {Llebaria},
  {Bout}, {Schwenn}, {Simnett}, {Bedford}, \& {Eyles}}]{brueckner95a}
{Brueckner}, G.~E., {Howard}, R.~A., {Koomen}, M.~J., {Korendyke}, C.~M.,
  {Michels}, D.~J., {Moses}, J.~D., {Socker}, D.~G., {Dere}, K.~P., {Lamy},
  P.~L., {Llebaria}, A., {Bout}, M.~V., {Schwenn}, R., {Simnett}, G.~M.,
  {Bedford}, D.~K., \& {Eyles}, C.~J. 1995, \solphys, 162, 357

\bibitem[{{Burkepile} {et~al.}(2004){Burkepile}, {Hundhausen}, {Stanger},
  {St.~Cyr}, \& {Seiden}}]{burkepile04a}
{Burkepile}, J.~T., {Hundhausen}, A.~J., {Stanger}, A.~L., {St.~Cyr}, O.~C., \&
  {Seiden}, J.~A. 2004, Journal of Geophysical Research (Space Physics), 109,
  3103

\bibitem[{{Delaboudiniere} {et~al.}(1995){Delaboudiniere}, {Artzner},
  {Brunaud}, {Gabriel}, {Hochedez}, {Millier}, {Song}, {Au}, {Dere}, {Howard},
  {Kreplin}, {Michels}, {Moses}, {Defise}, {Jamar}, {Rochus}, {Chauvineau},
  {Marioge}, {Catura}, {Lemen}, {Shing}, {Stern}, {Gurman}, {Neupert},
  {Maucherat}, {Clette}, {Cugnon}, \& {van Dessel}}]{delaboudiniere95a}
{Delaboudiniere}, J.-P., {Artzner}, G.~E., {Brunaud}, J., {Gabriel}, A.~H.,
  {Hochedez}, J.~F., {Millier}, F., {Song}, X.~Y., {Au}, B., {Dere}, K.~P.,
  {Howard}, R.~A., {Kreplin}, R., {Michels}, D.~J., {Moses}, J.~D., {Defise},
  J.~M., {Jamar}, C., {Rochus}, P., {Chauvineau}, J.~P., {Marioge}, J.~P.,
  {Catura}, R.~C., {Lemen}, J.~R., {Shing}, L., {Stern}, R.~A., {Gurman},
  J.~B., {Neupert}, W.~M., {Maucherat}, A., {Clette}, F., {Cugnon}, P., \& {van
  Dessel}, E.~L. 1995, \solphys, 162, 291

\bibitem[{{Demoulin} {et~al.}(1994){Demoulin}, {Henoux}, \&
  {Mandrini}}]{demoulin94a}
{Demoulin}, P., {Henoux}, J.~C., \& {Mandrini}, C.~H. 1994, \aap, 285, 1023

\bibitem[{{Forbes}(2000)}]{forbes00a}
{Forbes}, T.~G. 2000, \jgr, 105, 23153

\bibitem[{{Forbes} \& {Isenberg}(1991)}]{forbes91a}
{Forbes}, T.~G., \& {Isenberg}, P.~A. 1991, \apj, 373, 294

\bibitem[{{Gary}(1989)}]{gary89a}
{Gary}, G.~A. 1989, \apjs, 69, 323

\bibitem[{{Gary} \& {Moore}(2004)}]{gary04a}
{Gary}, G.~A., \& {Moore}, R.~L. 2004, \apj, 611, 545

\bibitem[{{Gopalswamy} {et~al.}(2004){Gopalswamy}, {Yashiro}, {Krucker},
  {Stenborg}, \& {Howard}}]{gopalswamy04a}
{Gopalswamy}, N., {Yashiro}, S., {Krucker}, S., {Stenborg}, G., \& {Howard},
  R.~A. 2004, Journal of Geophysical Research (Space Physics), 109, 12105

\bibitem[{{Greene}(1988)}]{greene88a}
{Greene}, J.~M. 1988, \jgr, 93, 8583

\bibitem[{{Handy} {et~al.}(1999){Handy}, {Acton}, {Kankelborg}, {Wolfson},
  {Akin}, {Bruner}, {Caravalho}, {Catura}, {Chevalier}, {Duncan}, {Edwards},
  {Feinstein}, {Freeland}, {Friedlaender}, {Hoffmann}, {Hurlburt}, {Jurcevich},
  {Katz}, {Kelly}, {Lemen}, {Levay}, {Lindgren}, {Mathur}, {Meyer}, {Morrison},
  {Morrison}, {Nightingale}, {Pope}, {Rehse}, {Schrijver}, {Shine}, {Shing},
  {Strong}, {Tarbell}, {Title}, {Torgerson}, {Golub}, {Bookbinder}, {Caldwell},
  {Cheimets}, {Davis}, {Deluca}, {McMullen}, {Warren}, {Amato}, {Fisher},
  {Maldonado}, \& {Parkinson}}]{handy99a}
{Handy}, B.~N., {Acton}, L.~W., {Kankelborg}, C.~C., {Wolfson}, C.~J., {Akin},
  D.~J., {Bruner}, M.~E., {Caravalho}, R., {Catura}, R.~C., {Chevalier}, R.,
  {Duncan}, D.~W., {Edwards}, C.~G., {Feinstein}, C.~N., {Freeland}, S.~L.,
  {Friedlaender}, F.~M., {Hoffmann}, C.~H., {Hurlburt}, N.~E., {Jurcevich},
  B.~K., {Katz}, N.~L., {Kelly}, G.~A., {Lemen}, J.~R., {Levay}, M.,
  {Lindgren}, R.~W., {Mathur}, D.~P., {Meyer}, S.~B., {Morrison}, S.~J.,
  {Morrison}, M.~D., {Nightingale}, R.~W., {Pope}, T.~P., {Rehse}, R.~A.,
  {Schrijver}, C.~J., {Shine}, R.~A., {Shing}, L., {Strong}, K.~T., {Tarbell},
  T.~D., {Title}, A.~M., {Torgerson}, D.~D., {Golub}, L., {Bookbinder}, J.~A.,
  {Caldwell}, D., {Cheimets}, P.~N., {Davis}, W.~N., {Deluca}, E.~E.,
  {McMullen}, R.~A., {Warren}, H.~P., {Amato}, D., {Fisher}, R., {Maldonado},
  H., \& {Parkinson}, C. 1999, \solphys, 187, 229

\bibitem[{{Hundhausen}(1997)}]{hundhausen97a}
{Hundhausen}, A.~J. 1997, in Coronal Mass Ejections: Geophysical Monograph 99,
  ed. N.~{Crooker}, J.~A. {Josely}, \& J.~{Feynman}

\bibitem[{{Klimchuk}(2001)}]{klimchuk01a}
{Klimchuk}, J.~A. 2001, Space Weather, 125, 143

\bibitem[{{Lara} {et~al.}(2006){Lara}, {Gopalswamy}, {Xie}, {Mendoza-Torres},
  {P{\'e}rez-Er{\'{\i}}quez}, \& {Michalek}}]{lara06a}
{Lara}, A., {Gopalswamy}, N., {Xie}, H., {Mendoza-Torres}, E.,
  {P{\'e}rez-Er{\'{\i}}quez}, R., \& {Michalek}, G. 2006, Journal of
  Geophysical Research (Space Physics), 111, 6107

\bibitem[{{Lau} \& {Finn}(1990)}]{lau90a}
{Lau}, Y.-T., \& {Finn}, J.~M. 1990, \apj, 350, 672

\bibitem[{{Li} {et~al.}(2006){Li}, {Schmieder}, {Aulanier}, \&
  {Berlicki}}]{li06a}
{Li}, H., {Schmieder}, B., {Aulanier}, G., \& {Berlicki}, A. 2006, \solphys, 18

\bibitem[{{Lin} {et~al.}(2001){Lin}, {Forbes}, \& {Isenberg}}]{lin01a}
{Lin}, J., {Forbes}, T.~G., \& {Isenberg}, P.~A. 2001, \jgr, 106, 25053

\bibitem[{{Lynch} {et~al.}(2004){Lynch}, {Antiochos}, {MacNeice}, {Zurbuchen},
  \& {Fisk}}]{lynch04a}
{Lynch}, B.~J., {Antiochos}, S.~K., {MacNeice}, P.~J., {Zurbuchen}, T.~H., \&
  {Fisk}, L.~A. 2004, \apj, 617, 589

\bibitem[{{Manoharan} \& {Kundu}(2003)}]{manoharan03a}
{Manoharan}, P.~K., \& {Kundu}, M.~R. 2003, \apj, 592, 597

\bibitem[{{Moon} {et~al.}(2002){Moon}, {Choe}, {Wang}, {Park}, {Gopalswamy},
  {Yang}, \& {Yashiro}}]{moon02a}
{Moon}, Y.-J., {Choe}, G.~S., {Wang}, H., {Park}, Y.~D., {Gopalswamy}, N.,
  {Yang}, G., \& {Yashiro}, S. 2002, \apj, 581, 694

\bibitem[{{Moore} \& {LaBonte}(1980)}]{moore80a}
{Moore}, R.~L., \& {LaBonte}, B.~J. 1980, Solar and Interplanetary Dynamics,
  207

\bibitem[{{Moore} {et~al.}(2001){Moore}, {Sterling}, {Hudson}, \&
  {Lemen}}]{moore01a}
{Moore}, R.~L., {Sterling}, A.~C., {Hudson}, H.~S., \& {Lemen}, J.~R. 2001,
  \apj, 552, 833

\bibitem[{{Priest} \& {Titov}(1996)}]{priest96a}
{Priest}, E.~R., \& {Titov}, V.~S. 1996, Phil. Trans. R. Soc. Lond. A, 354,
  2951

\bibitem[{{Scherrer} {et~al.}(1995){Scherrer}, {Bogart}, {Bush}, {Hoeksema},
  {Kosovichev}, {Schou}, {Rosenberg}, {Springer}, {Tarbell}, {Title},
  {Wolfson}, {Zayer}, \& {MDI Engineering Team}}]{scherrer95a}
{Scherrer}, P.~H., {Bogart}, R.~S., {Bush}, R.~I., {Hoeksema}, J.~T.,
  {Kosovichev}, A.~G., {Schou}, J., {Rosenberg}, W., {Springer}, L., {Tarbell},
  T.~D., {Title}, A., {Wolfson}, C.~J., {Zayer}, I., \& {MDI Engineering Team}.
  1995, \solphys, 162, 129

\bibitem[{{Schrijver} \& {Derosa}(2003)}]{schrijver03a}
{Schrijver}, C.~J., \& {Derosa}, M.~L. 2003, \solphys, 212, 165

\bibitem[{{Schrijver} {et~al.}(2005){Schrijver}, {DeRosa}, {Title}, \&
  {Metcalf}}]{schrijver05a}
{Schrijver}, C.~J., {DeRosa}, M.~L., {Title}, A.~M., \& {Metcalf}, T.~R. 2005,
  \apj, 628, 501

\bibitem[{{Sterling} \& {Moore}(2001)}]{sterling01a}
{Sterling}, A.~C., \& {Moore}, R.~L. 2001, \apj, 560, 1045

\bibitem[{{Sturrock}(1989)}]{sturrock89a}
{Sturrock}, P.~A. 1989, \solphys, 121, 387

\bibitem[{{T{\"o}r{\"o}k} \& {Kliem}(2005)}]{torok05a}
{T{\"o}r{\"o}k}, T., \& {Kliem}, B. 2005, \apjl, 630, L97

\bibitem[{{Vourlidas} {et~al.}(2002){Vourlidas}, {Buzasi}, {Howard}, \&
  {Esfandiari}}]{vourlidas02a}
{Vourlidas}, A., {Buzasi}, D., {Howard}, R.~A., \& {Esfandiari}, E. 2002, in
  ESA SP-506: Solar Variability: From Core to Outer Frontiers, ed. A.~{Wilson},
  91--94

\bibitem[{{Yurchyshyn} {et~al.}(2005){Yurchyshyn}, {Yashiro}, {Abramenko},
  {Wang}, \& {Gopalswamy}}]{yurchyshyn05a}
{Yurchyshyn}, V., {Yashiro}, S., {Abramenko}, V., {Wang}, H., \& {Gopalswamy},
  N. 2005, \apj, 619, 599

\bibitem[{{Zhang} \& {Low}(2005)}]{zhang05a}
{Zhang}, M., \& {Low}, B.~C. 2005, \araa, 43, 103

\bibitem[{{Zhou} {et~al.}(2006){Zhou}, {Wang}, \& {Zhang}}]{zhou06a}
{Zhou}, G.~P., {Wang}, J.~X., \& {Zhang}, J. 2006, \aap, 445, 1133

\end{thebibliography}
\end{document}